# Optimal width of quasi-crystalline slabs of dielectric cylinders to light transmission contrast

Ángel Andueza[1,2], Kang Wang[3], Jesús Pérez Conde[4] and Joaquín Sevilla[1,2]

[1] Dpto. Ing. Eléctrica y Electrónica Universidad Pública de Navarra

[2] Smart Cities Institute, Universidad Pública de Navarra

[3] Laboratoire de Physique des Solides, Laboratoire de Physique des Solides, UMR CNRS/Université Paris-Sud, 91405 Orsay, France

[4] Dpto. de Física Universidad Pública de Navarra

**Abstract**

Light confinement induced by resonant states in aperiodic photonic structures are interesting for many applications. A particular case of these resonances can be found in 2D quasi-crystalline arrangements of dielectric cylinders. These systems present a rather isotropic band gap as well as isolated in-gap photonic states (as a result of spatially localized resonances). These states are built by high symmetry polygonal clusters that can be regarded as photonic molecules. In this paper we study the transmission properties of a slab of glass cylinders arranged in approximants of decagonal quasi-crystalline structure. In particular, we investigate the influence of the slab width in the transmission contrast between the states and the gap. The study is both experimental and numerical in the microwave regime. We find that the best transmission contrast is found for a width of around 3 times the radiation wavelength. The transmission at the band gap region is mediated by the resonances of the photonic molecules. If the samples are thin enough they become transparent except around a resonance of the photonic molecule which reflects the incoming light. In this case the behavior is reminiscent of an "absorbing" molecule.

**Introduction**

Recent advances in the engineering of light confinement have demonstrated the ability to control the spectral properties of individual photonic modes in two dimensional disordered photonic structures, easing the creation of open transmission channels in strongly scattering media [1], [2]. The coupling of optical confinement modes offer the possibility to control light flow in random media as well as the construction of structures tuned to optimize light matter interaction. A lot of work has been recently done in the study of coupled resonator optical waveguides [3]–[6]. Quasi crystals are good candidates for these structures because they benefit from the properties of a quite isotropic band gaps as well as resonant structures imbedded in a perfectly determinist lattice [7]–[9]. Based on these structures interesting devices like subwavelength focusing lenses [10] or controllable light transmission channels [11], among others, have been built.

It has been proved that a quasi-crystalline arrangement of dielectric cylinders can present well defined transmission frequencies inside a wide bandgap [12], [13]. The stop band is mainly due to the global average order shown by the quasi-crystal structure, while the transmission frequencies are generated by certain highly symmetric structures of the quasi-crystal that present short range order and that resonate collectively at these definite frequencies. Making an analogy of the cylinders as photonic atoms we call these structures photonic molecules.

Previous works have explored the photonic states inside the gap of quasi crystal 2D structures generated in different ways. For example, in Ref. [14] individual cylinders are eliminated thereby introducing "atomic" defects. In Ref [15] they remove high-symmetry set of cylinders to create bigger resonators. An A alternative possibility to tailor the states in the gap is through the use of photonic molecules. In this paper we show how the same structures removed in [15] can act as photonic molecules, becoming a channel for the radiation to cross a slab of material.

A slab of dielectric cylinders could be a good photonic platform, suitable for different applications (filters, couplers, sensors, waveguides, etc.) because of the above-mentioned characteristics: an omnidirectional gap as well as the appearance of resonant states. In this paper we investigate the possible role of collective oscillators, or photonic molecules, as candidates to act as resonant states generators. Particularly we study the transmission of radiation through a cylinders based slab of decagonal symmetry quasi-crystals. The decagonal photonic molecules are at the slab center and we analyze how the transmission evolves with the slab width. At their resonance frequencies these collective oscillators, or photonic molecules, act as a bridge by coupling radiation form the input side of the slab to the output one. However, for very small slab widths, the effect is reversed and the resonance frequency of the molecules is absorbed in contrast with the rest of the -non resonant- spectrum. This behavior is reminiscent of the spectroscopic lines of actual molecules, which can be observed in the emission or the absorption spectra in different situations.

**Experimental**

The system is determined by a slab of cylinders, infinite in the plane perpendicular to the propagation direction, *z* direction in the following. The slab considered was made up with dielectric cylindrical rods placed at the points of a quasi-crystalline lattice (see figure 1(a)). We take as initial structure the Penrose QC tiling of decagonal symmetry and follow the process described by Wang [13]. We obtain an approximant structure that retains the same structure order at local scale than the original QC. We end up with a rectangular structure of 234 nodes with one of the substructures of decagonal symmetry placed in the center. This is convenient in order to investigate the effect of this structure as the width of the sample is changed.

The pattern of nodes is then transformed in circular holes drilled (by a numerical control machine) into two identical wooden pieces. Glass made cylinders, normally used in chemistry labs, were trimmed into pieces of 25 cm approximately. The glass rods were fixed in the wooden pieces as shown in figure 1(b). Rod diameter was $\Phi$=6 mm, and their dielectric permittivity of $\varepsilon$=4.5. The permittivity value was previously obtained from square and triangular lattices made up with the same rods. Then, we measured the transmission spectra of both regular lattices and fitted them with their respective simulations. Rod disposition was assessed by the common length of the two rhombus that compose the Penrose tiling, usually denoted by "*a*", and had a value of *a*= 12mm (see Figure 1(c)). In the definition of the coordinate axis, as indicated in Figure 1, the cylinders are aligned with the x axis, while z is the direction of propagation of the radiation.

The transmission spectra of the samples were measured with a network analyzer (HP 8722ES), spliced to rectangular horn antennas 79mm × 46mm (Narda model 639640). The antennas were aligned in the *z* direction separated by 200 mm approximately and facing each other, with the sample placed in between. These antennas show polarization sensitive coupling characteristics, with the electrical field parallel to the short side of its rectangular opening. We consider in this work the transverse magnetic (TM) polarization, with the electric field vector parallel to the cylinder axes. Spectra were registered differentially, with the free space transmission as the base line for lattice transmission measurements.

We have also performed numerical simulations of the same two dimensional pattern of cylinders but infinite in the y direction (rods infinitely long) and in the x direction. This was obtained by repeating the approximant with mirror unit cell boundary conditions in x and y axis. Simulations were performed with CST MICROWAVE STUDIO™, a commercial code based on the Finite Integration time-domain Method (FIM) [16].

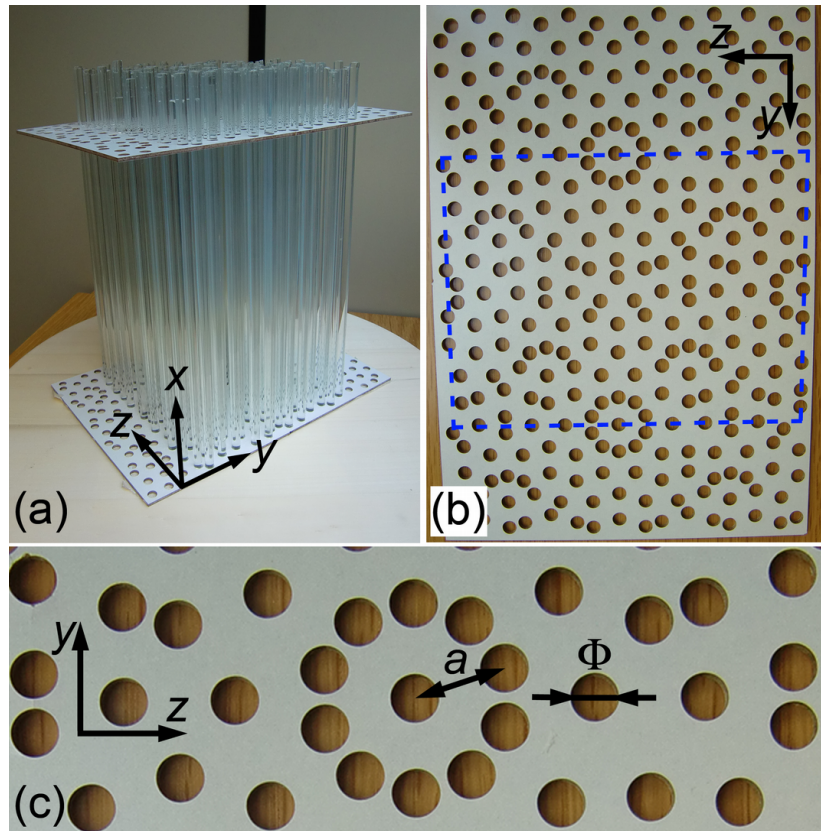

*Figure 1. The sample made by arranging glass cylinders in a decagonal quasi crystalline approximant lattice. (a) The sample completed with axis notation used in the text, (b) wooden piece with the pattern of holes. Quasi-crystalline approximant is represented by green dashed line. (c) Detail of the pattern indicating system parameters.*

**Results and discussion**

The spectra were calculated and measured, for different values of sample width starting with the complete sample (156.5 mm wide). Then the cylinders were removed row after row and measurements recorded for each case. The same situation was reproduced in the calculations. The results are presented in figure 2, with simulations on the left and the measurements on the right. Horizontal axis depicts the sample width, vertical the radiation frequency and the absorption value is represented through a color code, blue for low transmission and red for high transparency.

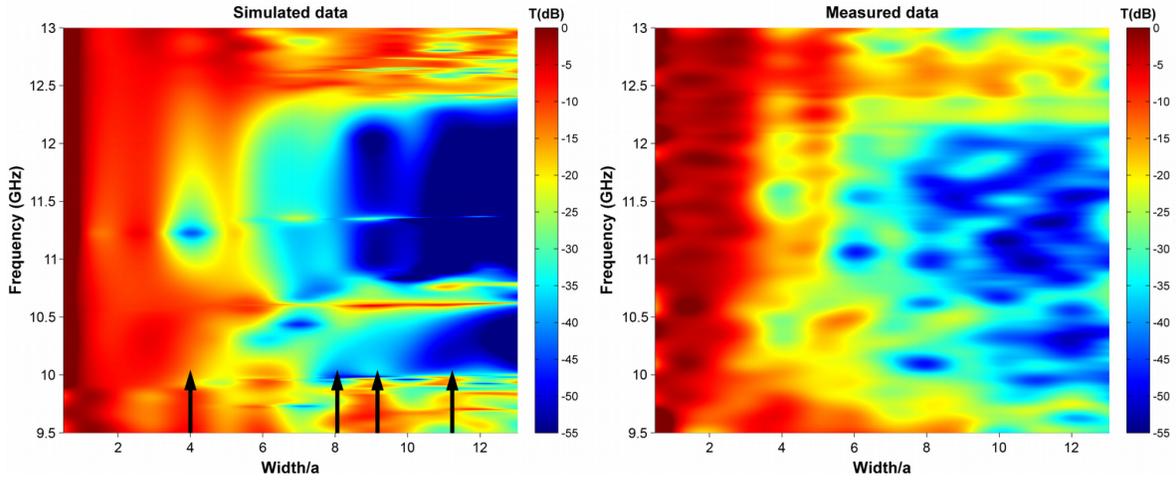

*Figure 2.- Color map plots of transmission spectra, simulated in the left and measured in the right, against sample width and frequency. The vertical axis depicts frequency while transmission value is color coded (red for full transmission, yellow for -20 dB and dark blue for -55 dB). Width is presented normalized to the lattice parameter a.*

Although the measurements present, as usual, a much noisier pattern, the results are consistent. In figure 3 we present the spectra, in a more standard representation, at four values of the sample width, marked with black arrows in figure 2. In this representation is easier to compare measured and simulated values. It can be seen that simulated data follow quite well the variations of the measured spectra.

The studied systems present a gap ranging from 10 to 12.5 GHz approximately. The gap is clearly seen (particularly in the simulated data) for sample widths larger than W/a = 6. As the sample gets thinner, the absorption decreases significantly and the radiation finds less impediment for crossing it. The frequency interval under consideration corresponds to wavelengths ranging from 23.1mm to 31.6 mm. If we normalize the wave length with the lattice parameter *a*, we get $\lambda/a = [1.9, 2.6]$ interval. So we can say that for sample widths below one radiation wavelength the sample is almost transparent, while for values over three wavelengths the gap is fully developed.

Inside the gap there are several states that can be easily spotted as lines of high transmission almost invariant with the frequency. The measured resonance around 11.6 GHz in figure 3 (a) is particularly intense. The starting hypothesis is that these light states could correspond to collective resonances in the highly symmetric areas of the decagonal structure placed at the center of our sample. These states have been extensively studied in [12], [13]. In order to check this idea we calculated the electric field distribution for different frequencies and widths of the sample. Four of these field distributions, which are representative of the behavior of the whole system, are plotted in figure 4.

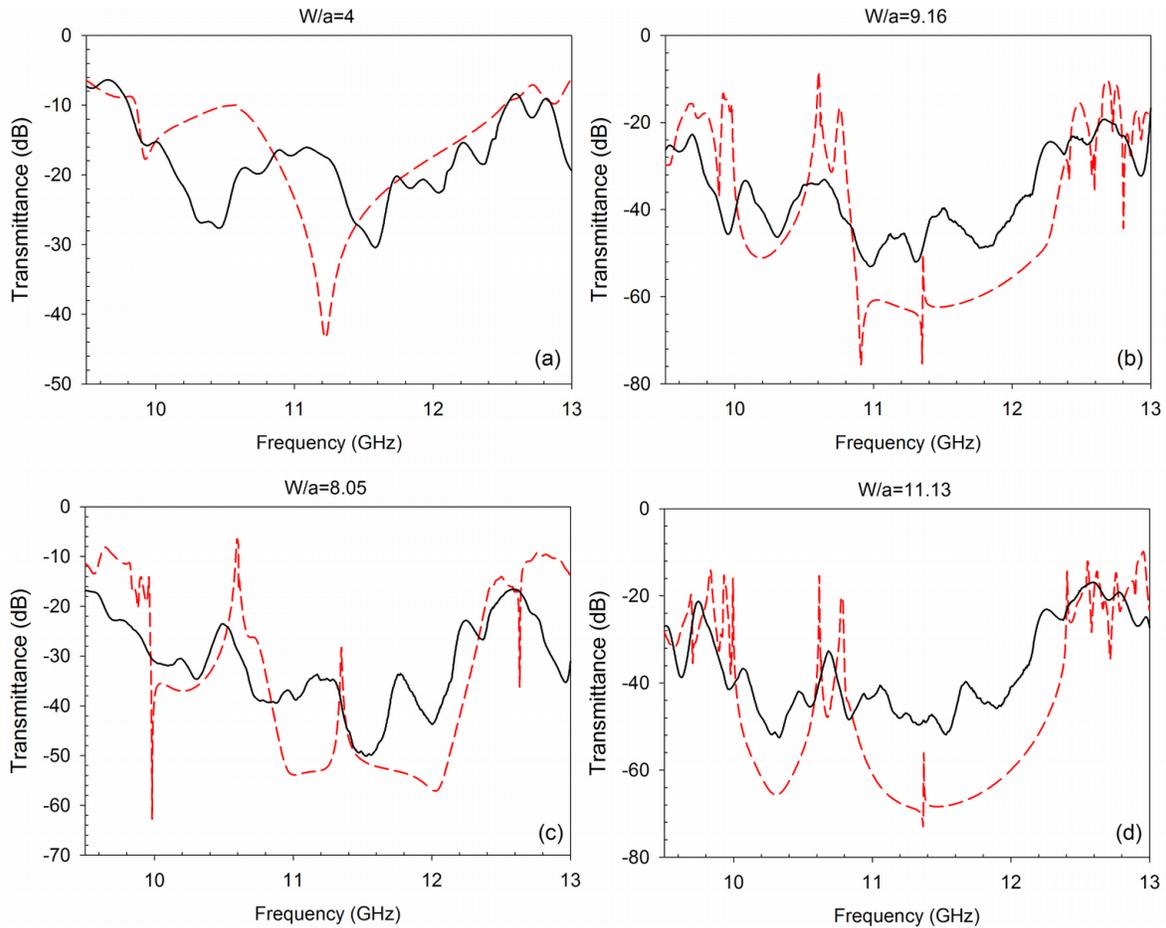

*Figure 3.- Transmission spectra of QC samples of different widths. Solid blue line correspond to experimental data and dashed ref one to simulations. These spectra correspond to the places marked with arrows in figure 2.*

For the cases of states inside the gap (B, C and D in figure 4) it is clear that the enhanced transmission at this particular frequencies is due to resonance modes of the decagonal symmetry ring at the center of the sample. These resonant modes are the lowest order states according to the classification of Wang [12], [13], as expected. Therefore we can conclude that the light transmission within the gap detected in our system is due to radiation coupling from the input side to the resonant localized patterns and from there to the output side.

It is interesting to note that even for much lower dielectric permittivity 13 in Ref. [12] versus 4.5 in our case, the resonances show the same field distributions and symmetries, and also their position inside the gap are similar. Therefore, what we see here is the possibility to build cylinder based "molecules", even with low permittivity rods as "photonic atoms".

Another interesting observation is the spot denoted by A in figure 4 (also present in the measured data). For a width of W/a= 4 and a frequency of 11.6 GHz there is an appreciable decrease in the observed transmission. It appears to be the reverse situation of the states inside the gap. Here, instead of an enhanced transmission for certain frequencies within the gap we have a decrease in the transmission band. If we analyze the electric field distribution at this point (A in figure 4) we see that the transmission blocking is caused also by resonant states of the photonic molecule previously considered. The state shows the same field distribution as the D-state inside the gap and appears at close frequency to the highly symmetric B resonance. All radiation that is not transmitted is reflected, giving rise to a standing wave as it interacts with the incident beam.

In a slight language abuse, we can extend the metaphor of the rod cluster as a photonic molecule with its states being its spectral lines. These spectral lines can be observed "in emission" and "in absorption". That is, these resonances increase the transmission in more opaque zones (spots B, D and D) or, on the contrary, reduce the transmitted intensity in the transparent regions (spot A). The emission and absorption (reflective, in fact) lines occur at different frequencies however, so the analogy is only partial.

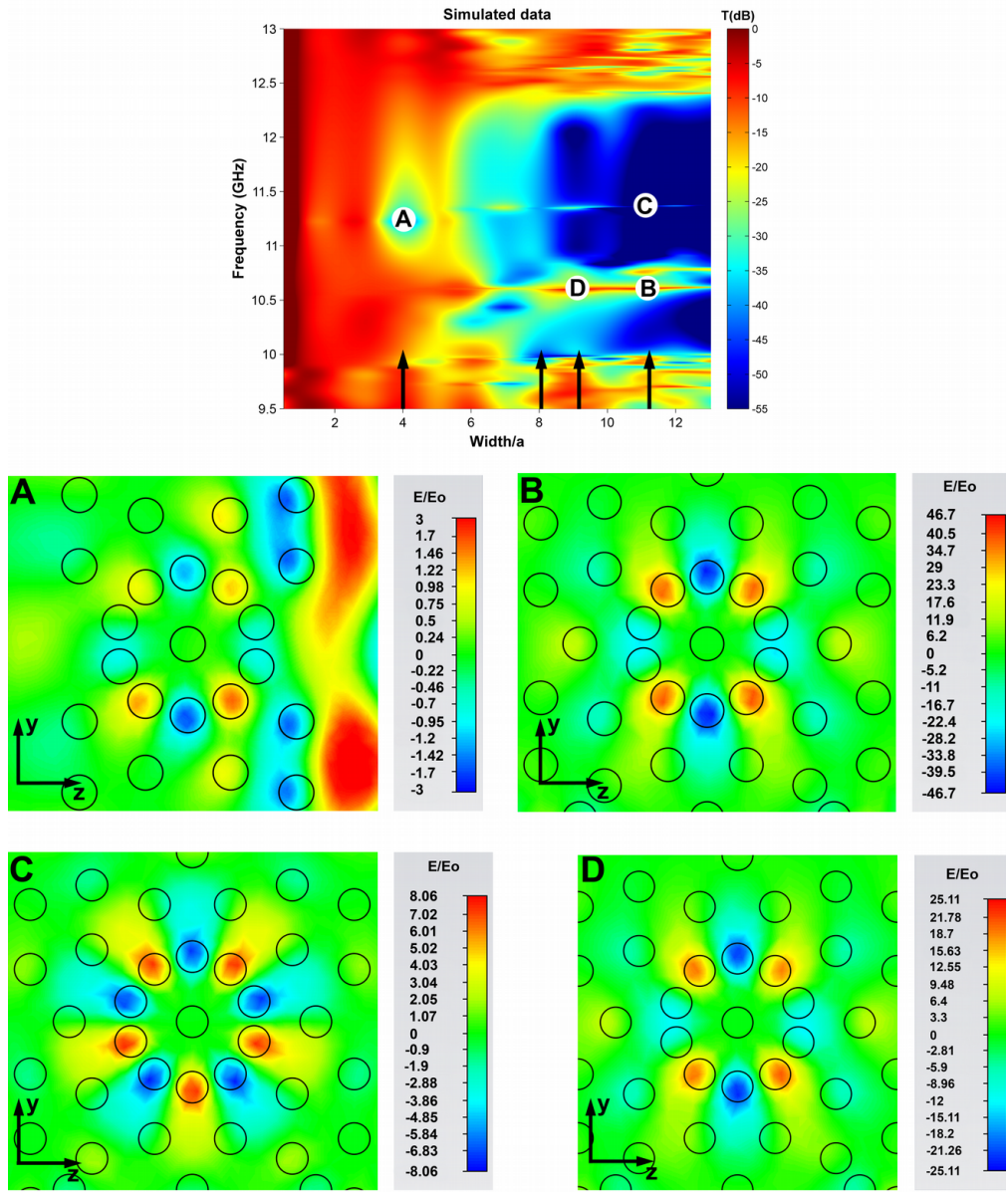

*Figure 4.- Electric field density calculated for samples of different widths at the frequencies of the transmission states as indicated over the color map. (a) 11.3 GHz, W=4 (b) 11.4 GHz, W= 11.13 (c) 10.8 GHz, W= 11.13 (d) 10.5 GHz, W= 9.16 .*

Finally, as we were interested in the transmission contrast between gap depth and resonance frequencies in the gap, we analyze their respective variation with the sample width. The gap depth grows with sample width almost linearly while at the frequencies of the resonant states the transmission remains at much higher values. This generates a difference in transmission that is maximum for intermediate values of sample depth. If the sample is too thin, the radiation can traverse it with few losses and the resonant states do not provide great contrast (all in red region in figure 2). On the other side, if the sample is too thick the gap is deep, but also the resonant states are nor efficient conveying the radiation form one side to the other (all in blue in figure 2). The midpoint where maximum contrast is attained is between W/a = 8 and W/a = 10 (spot D in figure 4, for example). This optimal width corresponds, approximately, to three wavelengths of the radiation of interest.

**Conclusions**

A sample of finite width of quasi-crystalline array of cylinders was numerically simulated and physically built and measured in the microwave range. The quasi-crystal was of decagonal symmetry with one of the high symmetry decagonal patterns placed in the middle of the sample. This decagonal pattern acts as a photonic molecule, holding resonant states that play a crucial role in the transmission behavior of the sample. Even for a dielectric glass cylinder permittivity as low as 4.5 (the value presented by our glass cylinders) the locally resonant states are similar to those described for much higher values. This is not obvious because of the very different efficiency behavior shown for both permittivity values.

The photonic molecule presents three states in the gap which show a significant contrast in transmission intensity with the surrounding frequency values. The optimal contrast is found when the sample width is around three times the wavelength of the incoming field, approximately.

The same states that enhance transmission for thick samples behave conversely, blocking it, in thin samples which are almost transparent at any frequency. This behavior can be regarded as similar to the spectroscopy pattern of usual molecules, where its states can be appear in emission as well as in absorption. We must realize, however, that there is no actual absorption but reflection in this case.

**References**

[1]    F. Riboli, N. Caselli, S. Vignolini, F. Intonti, K. Vynck, P. Barthelemy, A. Gerardino, L. Balet, L. H. Li, A. Fiore, M. Gurioli, and D. S. Wiersma, "Engineering of light confinement in strongly scattering disordered media.," *Nat. Mater.*, vol. 13, no. 7, pp. 720–5, Jul. 2014.

[2]  D. S. Wiersma, "Disordered photonics," *Nat. Photonics*, vol. 7, no. 3, pp. 188–196, Feb. 2013.

[3]  D. Valderamma and A. T. Reutov, "Propagation of light over irregular waveguiding structures formed that are formed by chains of coupled optic microresonators," *J. Commun. Technol. Electron.*, vol. 55, no. 3, pp. 316–324, Apr. 2010.

[4]  E. Ozbay, M. Bayindir, I. Bulu, and E. Cubukcu, "Investigation of localized coupled-cavity modes in two-dimensional photonic bandgap structures," *IEEE J. Quantum Electron.*, vol. 38, no. 7, pp. 837–843, Jul. 2002.

[5]  K. Guven and E. Ozbay, "Coupling and phase analysis of cavity structures in two-dimensional photonic crystals," *Phys. Rev. B*, vol. 71, no. 8, p. 085108, Feb. 2005.

[6]  S. Longhi, "Tunable dynamic Fano resonances in coupled-resonator optical waveguides," *Phys. Rev. A*, vol. 91, no. 6, p. 063809, Jun. 2015.

[7]  A. Della Villa, S. Enoch, G. Tayeb, F. Capolino, V. Pierro, and V. Galdi, "Localized modes in photonic quasicrystals with Penrose-type lattice," *Opt. Express*, vol. 14, no. 21, p. 10021, Oct. 2006.

[8]  A. Ricciardi, M. Pisco, A. Cutolo, A. Cusano, L. O'Faolain, T. F. Krauss, G. Castaldi, and V. Galdi, "Evidence of guided resonances in photonic quasicrystal slabs," *Phys. Rev. B*, vol. 84, no. 8, p. 085135, Aug. 2011.

[9]  W. Man, M. Florescu, K. Matsuyama, P. Yadak, G. Nahal, S. Hashemizad, E. Williamson, P. Steinhardt, S. Torquato, and P. Chaikin, "Photonic band gap in isotropic hyperuniform disordered solids with low dielectric contrast.," *Opt. Express*, vol. 21, no. 17, pp. 19972–81, Aug. 2013.

[10]  X. Zhang, W. Zhong, Z. Feng, Y. Wang, Z.-Y. Li, and D.-Z. Zhang, "Negative refraction and localized states of a classical wave in high-symmetry quasicrystals," *Philos. Mag.*, vol. 91, no. 19–21, pp. 2811–2819, Jul. 2011.

[11]  P. R. T, G. Zito, E. Di Gennaro, G. Abbate, and A. Andreone, "Control of the light transmission through a quasiperiodic waveguide.," *Opt. Express*, vol. 20, no. 23, pp. 26056–61, Nov. 2012.

[12]  K. Wang, "Light wave states in two-dimensional quasiperiodic media," *Phys. Rev. B*, vol. 73, no. 23, p. 235122, 2006.

[13]  K. Wang, "Structural effects on light wave behavior in quasiperiodic regular and decagonal Penrose-tiling dielectric media: A comparative study," *Phys. Rev. B*, vol. 76, no. 8, p. 085107, 2007.

[14]  M. Bayindir, E. Cubukcu, I. Bulu, and E. Ozbay, "Photonic band-gap effect, localization, and waveguiding in the two-dimensional Penrose lattice," *Phys. Rev. B*, vol. 63, no. 16, p. 161104, 2001.

[15]  C. Jin, B. Cheng, B. Man, Z. Li, and D. Zhang, "Two-dimensional dodecagonal and decagonal quasiperiodic photonic crystals in the microwave region," *Phys. Rev. B*, vol. 61, no. 16, pp. 10762–10767, 2000.

[16]  T. Weiland, in Computational Electromagnetics (Springer, 2003), pp. 183–198.